\documentclass[final,english]{bullsrsl}[2022/06/15]

\usepackage[latin1]{inputenc}
\usepackage[T1]{fontenc}
\usepackage{natbib} 
\usepackage{graphicx}

\begin{document}
\title{The 4m International Liquid Mirror Telescope project}

\author[affil={1}, corresponding]{Jean}{Surdej}
\author[affil={2,3}]{Bhavya}{Ailawadhi}
\author[affil={4,5}]{Talat}{Akhunov}
\author[affil={6}]{Ermanno}{Borra}
\author[affil={2,7}]{Monalisa}{Dubey}
\author[affil={2,7}]{Naveen}{Dukiya}
\author[affil={8}]{Jiuyang}{Fu}
\author[affil={8}]{Baldeep}{Grewal}
\author[affil={8}]{Paul}{Hickson}
\author[affil={2}]{Brajesh}{Kumar}
\author[affil={2}]{Kuntal}{Misra}
\author[affil={2,3}]{Vibhore}{Negi}
\author[affil={1}]{Anna}{Pospieszalska-Surdej}
\author[affil={2,9}]{Kumar}{Pranshu}
\author[affil={8}]{Ethen}{Sun}
\affiliation[1]{Institute of Astrophysics and Geophysics, University of Li\`{e}ge, All\'{e}e du 6 Ao$\hat{\rm u}$t 19c, 4000 Li\`{e}ge, Belgium}
\affiliation[2]{Aryabhatta Research Institute of observational sciencES (ARIES), Manora Peak, Nainital, 263001, India}
\affiliation[3]{Department of Physics, Deen Dayal Upadhyaya Gorakhpur University, Gorakhpur, 273009, India}
\affiliation[4]{National University of Uzbekistan, Department of Astronomy and Astrophysics, 100174 Tashkent, Uzbekistan}
\affiliation[5]{ Ulugh Beg Astronomical Institute of the Uzbek Academy of Sciences, Astronomicheskaya 33, 100052 Tashkent, Uzbekistan}
\affiliation[6]{Department of Physics, Universit\'{e} Laval, 2325, rue de l'Universit\'{e}, Qu\'{e}bec, G1V 0A6, Canada}
\affiliation[7]{Department of Applied Physics, Mahatma Jyotiba Phule Rohilkhand University, Bareilly, 243006, India}
\affiliation[8]{Department of Physics and Astronomy, University of British Columbia, 6224 Agricultural Road, Vancouver, BC V6T 1Z1, Canada}
\affiliation[9]{Department of Applied Optics and Photonics, University of Calcutta, Kolkata, 700106, India}


\correspondance{jsurdej@uliege.be}
\date{6th May 2023}
\maketitle


%

\begin{abstract}
The International Liquid Mirror Telescope (ILMT) project is a scientific collaboration in observational astrophysics between the Li{\`e}ge Institute of Astrophysics and Geophysics (Li{\`e}ge University, Belgium), the Aryabatta Research Institute of observational sciencES (ARIES, Nainital, India) and several Canadian universities (British Columbia, Laval, Montr{\'e}al, Toronto, Victoria and York). Meanwhile, several other institutes have joined the project: the Royal Observatory of Belgium, the National University of Uzbekistan and the Ulugh Beg Astronomical Institute (Uzbekistan) as well as the Pozna{\'n} Observatory (Poland). The Li{\`e}ge company AMOS (Advanced Mechanical and Optical Systems) has fabricated the telescope structure that has been erected on the ARIES site in Devasthal (Uttarakhand, India). It is the first liquid mirror telescope being dedicated to astronomical observations. First light was obtained on 29 April 2022 and commissioning is being conducted at the present time. In this short article, we describe and illustrate the main components of the ILMT. We also highlight the ILMT papers presented during the third BINA workshop, which discuss various aspects of the ILMT science programs.
\end{abstract}



\section{Introduction}

The ILMT is a 4-metre zenith-pointing liquid-mirror telescope located at Devasthal Peak in the Indian Himalayas. Taking advantage of the best seeing conditions and atmospheric absorption towards the zenith, it is being used to conduct a deep survey consisting of high S/N photometric and astrometric observations in the SDSS $g$, $r$ or $i$ spectral bands of a narrow strip of sky ($22^\prime$ in declination). In combination with an efficient 4k $\times$ 4k CCD camera and a unique prime-focus optical corrector, the images are obtained using the Time Delayed Integration (TDI) technique. The singly-scanned CCD frames correspond to an integration time of 102\,s, corresponding to the time the image of an object crosses the active area of the detector. The ILMT presently reaches V $\sim$ 22 mag (i-band) in a single scan but this limiting magnitude can be further improved by co-adding the nightly recorded images. 

The uniqueness of good cadence (one day) and deeper imaging with the ILMT make it possible to detect and characterize artificial satellites and space debris \citep{HicksonPProceedings}, solar system \citep{PospieszalskaSurdejAProceedings}, galactic \citep{GrewalBProceedings} and extra-galactic objects \citep{AkhunovTProceedings,SunEProceedings,KumarBProceedings}. The field covered by the ILMT during a full year is represented in equatorial, ecliptic and galactic coordinates in \citet{DubeyMProceedings}. The fast f/D $\sim$ 2.4 focal ratio of this telescope is particularly well adapted to the detection and characterization of low surface brightness objects \citep{FuJProceedings}. 

An image subtraction technique is also being applied to the nightly recorded observations in order to detect transients, objects exhibiting variations in flux or position \citep{PranshuKProceedings}.

ILMT data acquired during the fall of 2022 are being made freely available to the scientific community \citep{MisraKProceedings}. 

Following a campaign of observations that took place during the first commissioning period of the ILMT in October-November 2022, an astrometric and photometric pipeline has been developed to reduce and analyse the data \citep{AilawadhiBProceedings,DukiyaNProceedings,NegiVProceedings}. Some preliminary scientific results based upon these observations are presented in the different ILMT poster papers in these proceedings. 

After briefly reviewing the working principle of a liquid mirror, we describe and illustrate in the present paper the different components of the telescope.

\section{Working principle of the ILMT}

\begin{figure}[t]
\centering
\includegraphics[width=\textwidth]{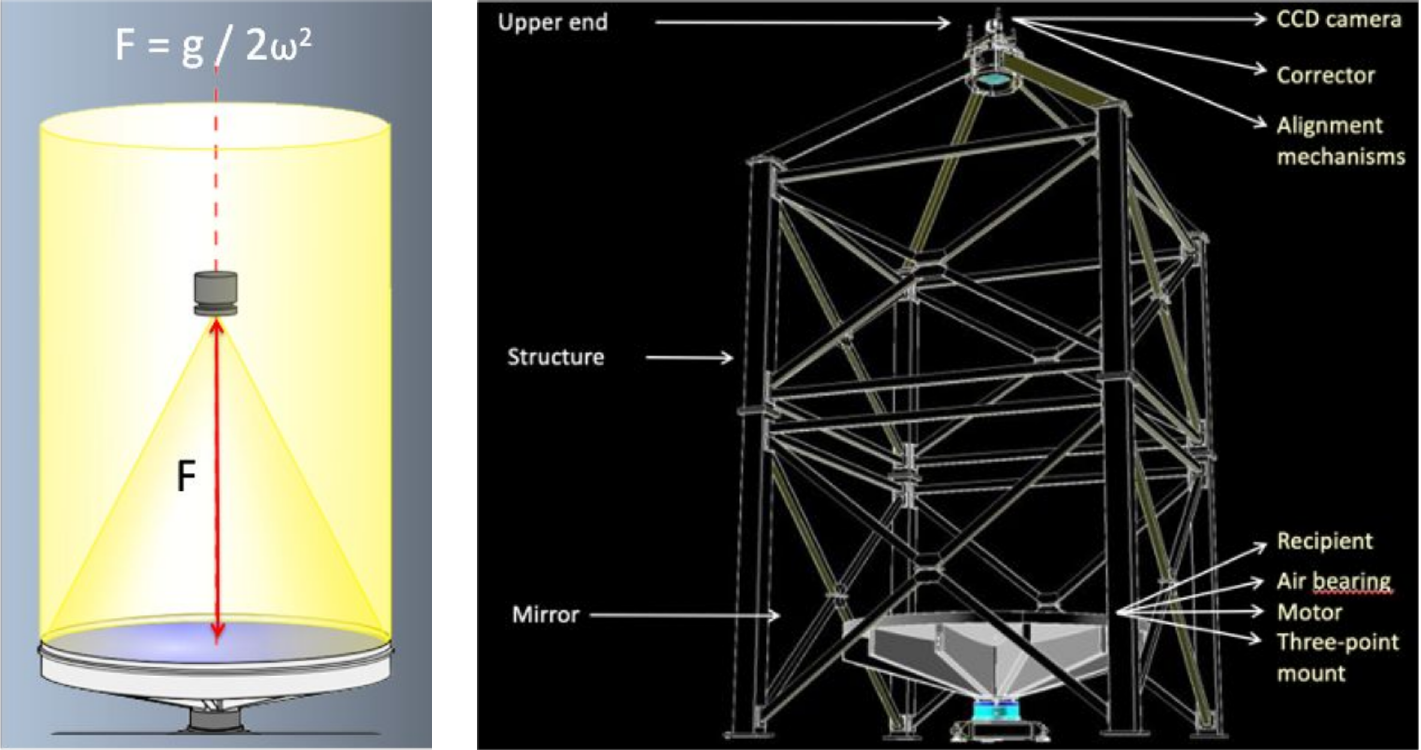}
\bigskip
\begin{minipage}{12cm}
  \caption{
    (Left) Mercury placed in a circular container spinning around the vertical axis focuses a beam of parallel light rays in the focal plane at a distance F from the vertex of the mirror (see text).
    (Right) The main components of a liquid mirror telescope: the mirror, the metallic structure and the upper end installed at the level of the prime focus.
    \label{Fig_1}}
\end{minipage}
\end{figure}

A perfect reflective paraboloid represents the ideal reference surface for an optical device to focus a beam of parallel light rays to a single point. This is how astronomical mirrors form images of distant stars in their focal plane. In this context, it is amazing that the surface of a liquid rotating around a vertical axis takes the shape of a paraboloid under the constant pull of gravity and centrifugal acceleration, the latter growing stronger at distances further from the central axis. The parabolic surface occurs because a liquid always sets its surface perpendicular to the net acceleration it experiences, which in this case is increasingly tilted and enhanced with distance from the central axis. The focal length F is proportional to the gravity acceleration g and inversely proportional to the square of the angular velocity $\omega$
(see Fig.\,\ref{Fig_1}, left).
In the case of the ILMT, the angular velocity $\omega$ is about 8 turns per minute, resulting in a focal length of about 8m. Given the action of the optical corrector, the effective focal length f of the D=4m telescope is about 9.44m, resulting in the widely open ratio f/D $\sim$ 2.4. In the case of the ILMT, a thin rotating layer of mercury naturally focuses the light from a distant star at its focal point located at $\sim$ 8 m just above the mirror, with the natural constraint that such a telescope always observes at the zenith.



Thanks to the rotation of the Earth, the telescope scans a strip of sky centred at a declination equal to the latitude of the observatory ($+29^\circ 21^\prime41.4^{\prime \prime}$ for the ARIES Devasthal observatory). The angular width of the strip is about $22^\prime$, a size limited by that of the detector (4k $\times$ 4k) used in the focal plane of the telescope. Since the ILMT observes the same region of the sky night after night, it is possible either to co-add the images taken on different nights in order to improve the limiting magnitude or to subtract images taken on different nights to make a variability study of the corresponding strip of sky. Consequently, the ILMT is very well-suited to perform variability studies of the strip of sky it observes. While the ILMT mirror is rotating, the linear speed at its rim is about 5.6 km/hr, i.e., the speed of a walking person.


\section{Technologies used for the ILMT}






The technology of liquid mirrors has been developed by the teams of Prof. E. Borra at Laval University and Prof. P. Hickson at the University of British Columbia (UBC, see \citealt{Hickson1994ApJ...436L.201H}). Liquid mirror Telescopes are relatively simple. Three components are required
(see Fig.\,\ref{Fig_1}, right):
a fixed metallic structure, an upper hand comprising a TDI optical corrector, a CCD camera and alignment mechanisms, and a mirror. The latter is made of a dish containing a reflecting metal liquid (mercury), an air bearing that supports the mirror, and a drive system. The choice of an air bearing system has been made in order to avoid as much as possible the transmission of vibrations from the turntable to the mercury. The air bearing is driven by a synchronous motor controlled by a variable-frequency AC power supply stabilised with a crystal oscillator. It has to support the weight of the mercury ($\sim$ 650 kg). To avoid transitory effects in the mercury, the rotation speed must be very stable. Relative variations in the rotation period must be smaller than typically one part in a million (\citealt{Surdej2006SPIE.6267E..04S}).



\begin{figure}[t]
\centering
\includegraphics[width=\textwidth]{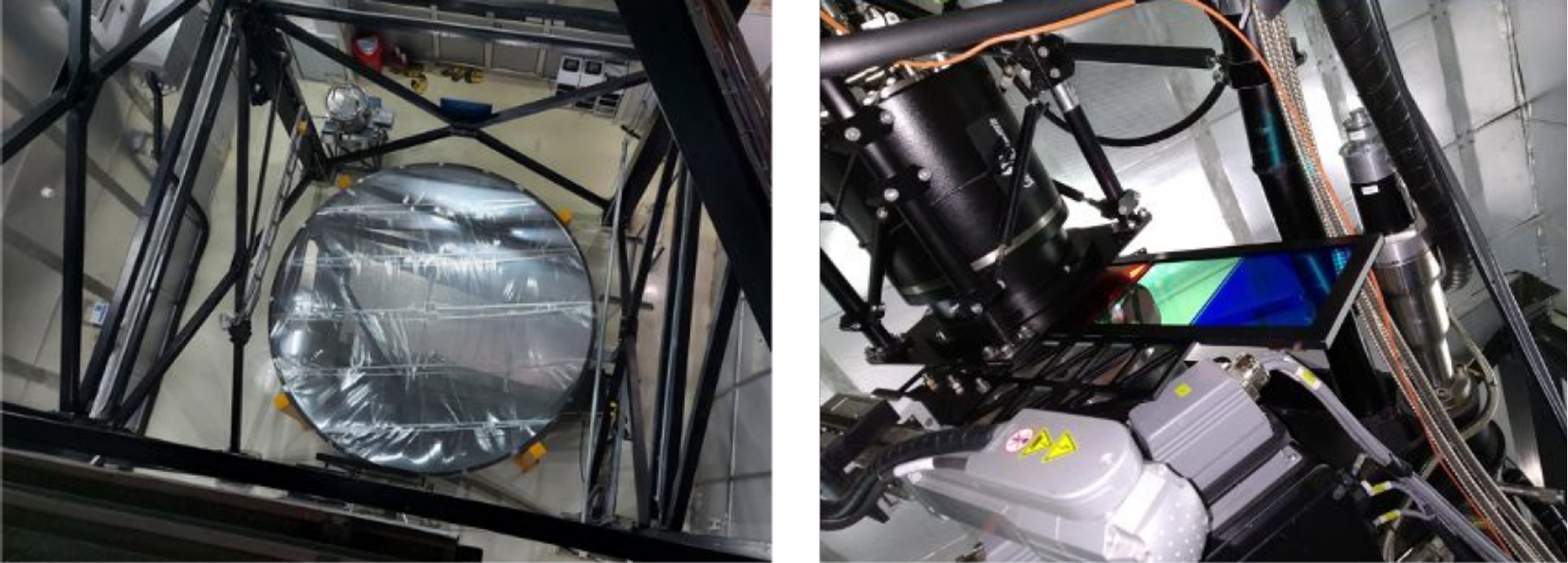}
\bigskip
\begin{minipage}{12cm}
  \caption{
    (Left) Top view of the ILMT structure and mirror filled with mercury and covered with mylar.
    (Right) The $g$, $r$ and $i$ Sloan filters installed in a tray just below the Spectral Instruments CCD camera.
    \label{Fig_2}}
\end{minipage}
\end{figure}

The left picture of Fig.\,\ref{Fig_2} shows a top view of the ILMT.
We see the mirror filled with mercury and covered with a very thin film of mylar in order to prevent any friction between the mercury and the ambient air that could produce ripples at its surface.
The right picture of Fig.\,\ref{Fig_2} shows the filter tray placed just under the CCD camera inside an interface structure at the prime focus.
This interface is equipped with mechanisms to focus the stellar images on the CCD and also to eventually translate and rotate the camera in the focal plane.


\section{Conclusions} 

Based upon ILMT observations that have been collected during 9 consecutive nights in October-November 2022 through the Sloan $g$, $r$ and $i$ spectral filters, composite CCD frames have been produced after correction for dark current, flat field, cosmic rays. They have been astrometrically and photometrically calibrated (\citealt{paper15}). Preliminary investigations have led to the following results: hundreds of transients and known asteroids have been identified and more than 80 satellite and space debris streaks have been detected. Very nice extended nebulae have also been imaged. 



\begin{acknowledgments}
The 4m International Liquid Mirror Telescope (ILMT) project results from a collaboration between the Institute of Astrophysics and Geophysics (University of Li\`{e}ge, Belgium), the Universities of British Columbia, Laval, Montreal, Toronto, Victoria and York University, and Aryabhatta Research Institute of observational sciencES (ARIES, India). The authors thank Hitesh Kumar, Himanshu Rawat, Khushal Singh and other observing staff for their assistance at the 4m ILMT.  The team acknowledges the contributions of ARIES's past and present scientific, engineering and administrative members in the realisation of the ILMT project. JS wishes to thank Service Public Wallonie, F.R.S.-FNRS (Belgium) and the University of Li\`{e}ge, Belgium for funding the construction of the ILMT. PH acknowledges financial support from the Natural Sciences and Engineering Research Council of Canada, RGPIN-2019-04369. PH and JS thank ARIES for hospitality during their visits to Devasthal. B.A. acknowledges the Council of Scientific $\&$ Industrial Research (CSIR) fellowship award (09/948(0005)/2020-EMR-I) for this work. M.D. acknowledges Innovation in Science Pursuit for Inspired Research (INSPIRE) fellowship award (DST/INSPIRE Fellowship/2020/IF200251) for this work. T.A. thanks Ministry of Higher Education, Science and Innovations of Uzbekistan (grant FZ-20200929344). This work is supported by the Belgo-Indian Network for Astronomy and astrophysics (BINA), approved by the International Division, Department of Science and Technology (DST, Govt. of India; DST/INT/BELG/P-09/2017) and the Belgian Federal Science Policy Office (BELSPO, Govt. of Belgium; BL/33/IN12).

\end{acknowledgments}

\begin{furtherinformation}

\begin{orcids}
\orcid{0000-0002-7005-1976}{Jean}{Surdej} 
\end{orcids}

\begin{authorcontributions}
This work results from a long-term collaboration to which all authors have made significant contributions.

\end{authorcontributions}

\begin{conflictsofinterest}
The authors declare no conflict of interest.
\end{conflictsofinterest}

\end{furtherinformation}

\bibliographystyle{bullsrsl-en}

\bibliography{S11-P12_SurdejJ}

\end{document}